\documentclass[twocolumn ,amssymb, amsmath,nobibnotes, aps, prl]{revtex4}
\usepackage{graphicx}
\usepackage{dcolumn}
\usepackage{bm}

\def\be{\begin{equation}}
\def\ee{\end{equation}}
\def\bea{\begin{eqnarray}}
\def\eea{\end{eqnarray}}

\begin{document}

\title{Background field method at finite temperature and density}
\author{M. Loewe}
\email{mloewe@fis.puc.cl} \affiliation{Facultad de F\'\i sica,
Pontificia Universidad Cat\'olica de Chile,\\ Casilla 306, Santiago
22, Chile.}
\author{S. Mendizabal}
\email{smendiza@fis.puc.cl} \affiliation{Facultad de F\'\i sica,
Pontificia Universidad Cat\'olica de Chile,\\ Casilla 306, Santiago
22, Chile.}
\author{J.C. Rojas}
\email{jurojas@ucn.cl} \affiliation{Departamento de F\'{\i}sica,
Universidad Cat\'{o}lica del Norte,\\ Casilla 1280, Antofagasta,
Chile}

\begin{abstract}
In this letter we make use of the Background Field Method (BFM) to
compute the effective potential of an $SU(2)$ gauge field theory, in
the presence of chemical potential and temperature. The main idea is
to consider the chemical potential as the background field. The
gauge fixing condition required by the BFM turns out to be exactly
the one we found in a previous article in a different context.
\end{abstract}



\maketitle

\section{Introduction}

The background field method (BFM) is an easy and common tool for
quantizing gauge fields without breaking explicitly the gauge
invariance. This method, well described in \cite{abbott, pascual,
weinberg}, can be used to find, perturbatively, in a simple way, the
effective action \cite{abbott2}. The BRST, the Slavnov-Taylor and
the Ward identities are preserved \cite{ferrari}. It is also
possible to show, in this frame, the renormalization of the standard
model \cite{denner}.

The extension of the BFM to theories at finite temperature, and/or
densities, has not been properly formulated. For example, there are
ambiguities in the formulation of the renormalization group, so
that, finally there is not a unique answer for the thermal/density
behavior of the running coupling constants \cite{chai}. The first
attempt to extend the BFM to finite temperature was proposed in
\cite{chai1} employing the thermal renormalization group
\cite{matsumoto}.

Recently we have discussed how to compute the thermodynamical
potential ($\Omega$) of the standard model, in the presence of
finite chemical potentials and temperature, using a new gauge fixing
condition that allows to separate the contribution of the different
fields \cite{lmr1}. This gauge fixing condition can be interpreted
as an extension of the well known $R_{\xi}$ gauge introduced by 't
Hooft. In this letter, we show that this gauge fixing condition
emerges in a natural way from a description based on the BFM, by
interpreting the chemical potentials as background fields. Notice
that this is a new way of handling chemical potentials, since in the
BFM they are not introduced as Lagrange multipliers associated to
conserved charges. An advantage of this way of handling chemical
potentials is the fact it is not necessary to compute the conserved
charges, and to integrate over the canonical field momenta, as it is
usually done, see for example \cite{kapusta2}.

We will concentrate our discussion on a pure $SU(2)$ Yang-Mills
theory, considering, afterwards, the inclusion of scalar and fermion
fields. We show how to compute the effective potential according to
the BFM prescription, in the presence of chemical potential and
temperature.


\section{Pure $SU(2)$ gauge theory}

As it is well known, the generating functional for a non Abelian
gauge theory is given by

\begin{eqnarray}
Z[J]&=&\int \textrm{D}A\det\left[\frac{\delta G^{a}}{\delta
w^{b}}\right]\exp i\left[S[A]\right.\nonumber\\
&&-\left.\frac{1}{2\xi}G\cdot G+J\cdot A\right],
\end{eqnarray}

\noindent where $A_{\mu}^{a}$ is the $SU(2)$ gauge field, $G^a$ is
the gauge fixing condition, $w^b$ are the infinitesimal gauge
parameters and the classical action $S[A]$ is given by

\begin{equation}
S=-\frac{1}{4}\int{\textrm{d}^{4}xF^{a}_{\mu\nu}F^{\mu\nu}_{a}},
\end{equation}

\noindent where

\begin{equation}
F^{a}_{\mu\nu}=\partial_{\mu}A^{a}_{\nu}-\partial_{\nu}A^{a}_{\mu}+g\epsilon^{abc}A^{b}_{\mu}A^{c}_{\nu}.
\end{equation}

According to the BFM, we must shift our field
$A_{\mu}^{a}\rightarrow A_{\mu}^{a}+B_{\mu}^{a}$, where
$B_{\mu}^{a}$ is the background field. Then, the new generating
functional will be

\begin{eqnarray}
\tilde{Z}[J,B]&=&\int
\textrm{D}A\det\left[\frac{\delta\tilde{G}^{a}}{\delta
w^{b}}\right]\exp i{\Large[} S[A+B]\nonumber\\
&&-\left.\frac{1}{2\xi}\tilde{G}\cdot\tilde{G}+J\cdot A\right],
\label{zeta}
\end{eqnarray}

\noindent where $\tilde{G}^a$ is the modified gauge fixing
condition. In order to maintain gauge symmetry, we need a
$\tilde{G}^a$ of the following shape

\begin{equation}
\tilde{G}^{a}=\partial^{\mu}A^{a}_{\mu}+g\epsilon^{abc}B^{b}_{\mu}A^{\mu}_{c}\equiv\bar{D}_{\mu}A^{\mu}_{a}.
\end{equation}

\noindent The new infinitesimal transformations, will be defined
according to

\begin{eqnarray}\label{transfmormation}
\delta A^{a}_{\mu}&\equiv&-\epsilon^{abc}w^{b}A^{c}_{\mu}\nonumber\\
\delta
B^{a}_{\mu}&\equiv&-\epsilon^{abc}w^{b}B^{c}_{\mu}+\frac{1}{g}\partial_{\mu}w^{a}\nonumber\\
\delta J^{a}_{\mu}&\equiv&-\epsilon^{abc}w^{b}J^{c}_{\mu},
\end{eqnarray}

\noindent so that the gauge symmetry remains unbroken, including the
$\tilde{G}\cdot\tilde{G}$ term. Notice that the gauge field
transforms as a matter field and the gauge parameters appear
associated to the background field \cite{weinberg}. The sum of both
fields transforms in the usual way

\begin{eqnarray}
\delta(A^{a}_{\mu}+B^{a}_{\mu})=-\epsilon^{abc}w^{b}(A^{c}_{\mu}+B^{c}_{\mu})+\frac{1}{g}\partial_{\mu}w^{a},
\end{eqnarray}

\noindent i.e. as gauge fields, so that the classical action in
(\ref{zeta}) remains invariant. In the loop calculation of the
effective action, the background field ($B^{a}_{\mu}$) appears as
external amputated legs, whereas the quantum gauge fields
($A^{a}_{\mu}$) and the ghost fields ($\eta^{a}$) live only in
internal lines.

The modified Lagrangian, including the ghost fields will read

\begin{equation}
\mathcal{L}_{mod}=\mathcal{L}_{A_{\mu}}+\mathcal{L}_{ghost}+\mathcal{L}_{GF},
\end{equation}

\noindent such that

\begin{eqnarray}
\mathcal{L}_{A_{\mu}}&=&-\frac{1}{4}\left(B^{a}_{\mu\nu}+\bar{D}_{\mu}A^{a}_{\nu}-\bar{D}_{\nu}A^{a}_{\mu}\right.
\left.+g\epsilon^{abc}A^{b}_{\mu}A^{c}_{\nu}\right)^{2},\nonumber\\
\mathcal{L}_{ghost}&=&-(\bar{D}_{\mu}\eta^{*}_{a})(\bar{D}^{\mu}\eta_{a}-g\epsilon^{abc}\eta_{b}A^{\mu}_{c}),\nonumber\\
\mathcal{L}_{GF}&=&-\frac{1}{2\xi}(\bar{D}_{\mu}A^{\mu}_{a})^2,
\end{eqnarray}

\noindent where

\begin{eqnarray}
\bar{D}_{\mu}A^{a}_{\nu}&\equiv&\partial_{\mu}A^{a}_{\nu}+g\epsilon^{abc}B^{b}_{\mu}A^{c}_{\nu},\\
\bar{D}_{\mu}\eta^{a}&\equiv&\partial_{\mu}\eta^{a}+g\epsilon^{abc}B^{b}_{\mu}\eta^{c},
\end{eqnarray}

\noindent and

\begin{equation}
B^a_{\mu\nu}\equiv\partial_{\mu}B^a_{\nu}-\partial_{\nu}B^a_{\mu}+g\epsilon^{abc}B^b_{\mu}B^c_{\nu}.
\end{equation}


\section{Chemical potentials as background fields}

The basic idea of this section is to introduce the chemical
potential ($\mu$) as the background field. This is different from
the usual approach, where chemical potentials appear in covariant
derivatives as constant external time component gauge fields. We
will use the following prescription

\begin{equation}\label{cambio}
B^{a}_{\mu}=\frac{\mu}{g}v_{\mu}\delta^{a3},
\end{equation}

\noindent where $v_{\mu}$ is a 4-velocity with respect to the
thermal bath, that allows us to keep a formal covariant language,
although, finally we have to choose the frame of reference where the
heat bath is at rest, i.e. $v_{\mu}=(1,0,0,0)$. We have chosen the
third component of the internal $SU(2)$ group. This restriction
corresponds to a simple orientation in the group manifold. The
appearance of the quotient $\mu/g$ is a consequence of keeping the
usual BFM relation between $Z_{B}$ and $Z_{g}$, the background field
and coupling constant renormalization factors, respectively, given
by

\begin{equation}
Z_{g}=Z_{B}^{-1/2}
\end{equation}

The gauge fixing conditions acquires the form

\begin{equation}
G^a=\partial_{\mu}A^{a}_{\mu}+\mu\epsilon^{abc}A^{c}_{\mu}v^{\mu}\delta^{3b}.
\end{equation}

We would like to emphasize that the same gauge fixing condition was
found in our previous paper \cite{lmr1}, using a complete different
approach. In \cite{lmr1}, we found an exact expression for the
effective potential of the Weinberg-Salam model in the presence of
chemical potentials and thermal effects. The idea was to diagonalize
the effective potential to get separate contributions from each
field. The splitting of the effective potential is not possible
without our gauge fixing condition. For example, in \cite{kapusta}
the author gives an expression for the effective potential, but only
in the high temperature expansion.


To show the efficiency of this method, we will proceed with the
calculation in the one loop approximation of the effective potential
for a pure gauge theory. As it is well known the one-loop thermal
effective action is given by

\begin{eqnarray}\label{effect}
\exp \Gamma^{\beta}_{1}[\phi_{c}]=\int \textrm{D}[Fields]\exp
\int_{0}^{\beta}\textrm{d}\tau\int
\textrm{d}^3x\mathcal{L}_{q}(\bar{x}),
\end{eqnarray}

\noindent where we shifted to Euclidean metric ($\tau=it$), with

\begin{equation}
\bar{x}=(-i\tau,{\bf x}), \qquad \bar{p}=(i\omega_{n},{\bf p}),
\end{equation}

\noindent with $\omega_{n}=2\pi n/\beta$ for bosons and
$\omega_{n}=2\pi(n+1)/\beta$ for fermions. In equation
(\ref{effect}) the $\mathcal{L}_{q}$ denotes the quadratic
Lagrangian for the $A^{a}_{\mu}$ and $\eta^{a}$ fields. Now, as it
is well known, the effective potential corresponds to the effective
action, by taking the classical field as a constant.

\begin{equation}
\Gamma^{\beta}_{1}[\phi_{c}=constant]=-\beta\int \textrm{d}^3x
\hspace{0.1 cm}\Omega^{\beta}_{eff}.
\end{equation}

Since the internal lines in the Feynman diagrams are associated to
the gauge and the ghosts fields, we will need to find the quadratic
Lagrangian in both fields, in order to form

\begin{eqnarray}
\int_{0}^{\beta}\hspace{-0.1 cm}{\textrm{d}\tau}\hspace{-0.1 cm}\int
\textrm{d}^3 x\mathcal{L}_{q}=\hspace{-0.1 cm}&-&\hspace{-0.1
cm}\frac{1}{2}\int \textrm{d}\bar{x}'\hspace{-0.1
cm}\int\hspace{-0.1 cm}
\textrm{d}\bar{x}A^a_{\mu}(\bar{x}')B^{\mu\nu}_{ab}(\bar{x}',\bar{x})A^b_{\nu}(\bar{x})\nonumber\\
&-&\int \textrm{d}\bar{x}'\int
\textrm{d}\bar{x}\omega^{*}(\bar{x}')C(\bar{x}',\bar{x})\omega(\bar{x}),
\end{eqnarray}

Now, the effective potential will be given by

\begin{equation}
-\beta\int \textrm{d}^3x \hspace{0.1
cm}\Omega_{eff}=-\frac{1}{2}\textrm{Tr}\ln\textbf{B}+\textrm{Tr}\ln\textbf{C}.
\end{equation}

\noindent the quadratic Lagrangian for a pure gauge theory is

\begin{eqnarray}
\mathcal{L}_{q}&=&-\frac{1}{4}(\bar{D}_{\mu}A^{a}_{\nu}-\bar{D}_{\nu}A^{a}_{\mu})^{2}-\frac{1}{2}B^{\mu\nu}_{a}\epsilon^{abc}A^{b}_{\mu}A^{c}_{\nu}\nonumber\\
&&-\frac{1}{2\xi}(\bar{D}_{\mu}A^{\mu}_{a})^{2}-(\bar{D}_{\mu}\omega^{*}_{a})(\bar{D}^{\mu}\omega_{a}).
\end{eqnarray}

Taking into account the choice (\ref{cambio}), we notice that $
B_{\mu\nu}^{a}=0$. Since $Z_B^{1/2}$ multiplies $ B_{\mu\nu}^{a}$,
we need three different chemical potentials $\mu^a$ associated to a
particular flavor. Otherwise $ B_{\mu\nu}^{a}$ vanishes and we are
not able to carry on the renormalization procedure \cite{weinberg}.
The idea is that renormalizability has already been proved, and then
we are free to select one direction in the isospin space in order to
compute the effective potential. This remind us the problem that
appears when quantizing gauge field theories, between the $R_{\xi}$
and the unitary gauge.

A very simple way
to calculate the effective potential is to write each field
explicitly. Let us choose, to simplify the calculations, $\xi=1$.
The sum of all contributions of the gauge fields will give

\begin{eqnarray}\label{amu}
\mathcal{L}_{q}&=&-\frac{1}{2}[\partial_{\mu}A^{1}_{\nu}\partial^{\mu}A^{\nu}_{1}+\mu^{2}A^{1}_{\nu}A^{\nu}_{1}]\nonumber\\
&&-\frac{1}{2}[\partial_{\mu}A^{2}_{\nu}\partial^{\mu}A^{\nu}_{2}+\mu^{2}A^{2}_{\nu}A^{\nu}_{2}]\nonumber\\
&&-\frac{1}{2}[\partial_{\mu}A^{3}_{\nu}\partial^{\mu}A^{\nu}_{3}]\nonumber\\
&&-\mu
v_{\mu}(A^{\nu}_{1}\partial^{\mu}A^{2}_{\nu}-A^{\nu}_{2}\partial^{\mu}A^{1}_{\nu}).
\end{eqnarray}

As usual, we will write this lagrangian in Euclidean metric, and the
calculation of the effective potential will be given by taking the
traces of (\ref{amu}). After evaluating the sums over the Matsubara
frequencies, we find the gauge field contribution to the thermal
effective potential

\begin{widetext}

\begin{equation}
\Omega^{\beta}_{A_{\mu}}=\frac{1}{2\beta}\int{d^{3}k\left(8\ln{\left[(1-e^{-\beta(|{\bf
k}|+\mu)})(1-e^{-\beta(|{\bf k}|-\mu)})\right]}+8\ln{(1-e^{-\beta
|{\bf k}|})}\right)}.
\end{equation}

Although it seems that we have extra degrees of freedom, we must not
forget that we still need to calculate the contribution of the ghost
fields, which reads

\begin{equation}
\Omega^{\beta}_{ghosts}=-\frac{1}{\beta}\int{d^{3}k\left(2\ln{\left[(1-e^{-\beta(|{\bf
k}|+\mu)})(1-e^{-\beta(|{\bf k}|-\mu)})\right]}+2\ln{(1-e^{-\beta
|{\bf k}|})}\right)}.
\end{equation}

So, the final result for the effective potential is

\begin{equation}
\Omega^{\beta}_{eff}=\frac{1}{\beta}\int{d^{3}k\left(2\ln{\left[(1-e^{-\beta(|{\bf
k}|+\mu)})(1-e^{-\beta(|{\bf k}|-\mu)})\right]}+2\ln{(1-e^{-\beta
|{\bf k}|})}\right)}.
\end{equation}

\end{widetext}

This is exactly the result we wanted to obtain. We can see that we a
have massless and chargeless gauge boson with two degrees of
freedom, and two massless and charged gauge bosons with two degrees
of freedom each.

\section{Scalar and fermion $SU(2)$ gauge symmetry}

First we will concentrate our attention on a scalar $SU(2)$ gauge
invariant theory, given by the following Lagrangian

\begin{equation}
\mathcal{L}=\frac{1}{2}(D_{\mu}\phi^a)^{T}(D^{\mu}\phi^a)-\frac{1}{4}F^{a}_{\mu\nu}F^{\mu\nu}_{a}-V(\phi),
\end{equation}

\noindent where $\phi$ belongs to the adjoint representation, i.e.
it is given by a real scalar triplet

\begin{eqnarray}
\phi=\left(
\begin{array}{c}
\phi_{1}\\
\phi_{2}\\
\phi_{3}
\end{array}
\right),
\end{eqnarray}

\noindent and

\begin{equation}
D_{\mu}\phi^a=(\partial_{\mu}-igA_{\mu})\phi^a,
\end{equation}

\begin{equation}
F^{a}_{\mu\nu}=\partial_{\mu}A^{a}_{\nu}-\partial_{\nu}A^{a}_{\mu}+g\epsilon^{abc}A^{b}_{\mu}A^{c}_{\nu},
\end{equation}

\noindent with

\begin{equation}
A_{\mu}=A^{b}_{\mu}T^{b}
\end{equation}

The classical potential is given by

\begin{equation}
V(\phi)=\frac{m^{2}}{2}\phi^{T}\phi+\frac{\lambda}{4}(\phi^{T}\phi)^{2},
\end{equation}

In the $SU(2)$ case, the group generators will be given by
$T^{a}=\omega^{a}/2$, where

\begin{eqnarray}
\omega_{1}&=&\left(
\begin{array}{ccc}
0 & 0 & 0\\
0 & 0 & -i\\
0 & i & 0
\end{array}
\right),\qquad
 \omega_{2}=\left(
\begin{array}{ccc}
0 & 0 & i\\
0 & 0 & 0\\
-i & 0 & 0
\end{array}
\right),\\
 \omega_{3}&=&\left(
\begin{array}{ccc}
0 & -i & 0\\
i & 0 & 0\\
0 & 0 & 0
\end{array}
\right),
\end{eqnarray}

Now, following the BFM prescription, we will expand our fields in
the following way

\begin{eqnarray}
\phi^a & \rightarrow & \phi^a+\bar{\phi}^a,\nonumber\\
A^{a}_{\mu} & \rightarrow & A^{a}_{\mu}+B^{a}_{\mu},
\end{eqnarray}

\noindent where

\begin{eqnarray}
\bar{\phi}^{a}=\left(
\begin{array}{c}
0\\
0\\
\nu
\end{array}
\right),
\end{eqnarray}

\noindent and

\begin{equation}
B^{a}_{\mu}=\frac{\mu}{g}v_{\mu}\delta^{a3}.
\end{equation}

\noindent Here $\bar{\phi}^a$ is the vacuum expectation value of the
scalar fields and $B^{a}_{\mu}$ is the background field associated
to the chemical potential. These two background fields are constant.
The Lagrangian will now read

\begin{eqnarray}
\mathcal{L}=&&\frac{1}{2}[\tilde{D}_{\mu}(\phi^a+\bar{\phi}^a)]^{T}[\tilde{D}^{\mu}(\phi^a+\bar{\phi}^a)]-V(\phi)\nonumber\\&-&
\frac{1}{4}(\bar{D}_{\mu}A^{a}_{\nu}-\bar{D}_{\nu}A^{a}_{\mu}+g\epsilon^{abc}A_{\mu}^{b}A_{\nu}^{c})^{2},
\end{eqnarray}

\noindent with

\begin{eqnarray}
\tilde{D}_{\mu}\phi^a & = & (\partial_{\mu}-igT^{b}(A^{b}_{\mu}+B^{b}_{\mu}))\phi^a,\nonumber\\
\bar{D}_{\mu}A_{a}^{\mu} & = &
\partial_{\mu}A^{\mu}_{a}+g\epsilon^{abc}B_{\mu}^{b}A^{\mu}_{c}.
\end{eqnarray}

The gauge fixing condition should be treated in the same way as
before, but this time incorporating the scalar fields

\begin{equation}
\mathcal{L}_{GF}=-\frac{1}{2\xi}[\bar{D}_{\mu}A^{\mu}_{a}-ig\xi\phi^{T}T^{a}\bar{\phi}]^{2}.
\end{equation}

Since

\begin{equation}
[T^{a},T^{b}]=i\epsilon^{abc}T^{c},
\end{equation}

\noindent and the Goldstone theorem says

\begin{eqnarray}
iT^{a}\bar{\phi}&=&0, \qquad \textrm{for every unbroken
symmetry},\nonumber\\
iT^{a}\bar{\phi}&\neq&0, \qquad \textrm{for every broken symmetry},
\end{eqnarray}

\noindent we can see that this gauge fixing condition will remove
every quadratic mixing between the scalar and the gauge fields.

For the calculation of the effective potential we only need the
quadratic terms in the Lagrangian,
 $\mathcal{L}^{q}=\mathcal{L}^{q}_{\phi}+\mathcal{L}^{q}_{A_{\mu}}+\mathcal{L}^{q}_{GF}+\mathcal{L}^{q}_{ghost}$, with

\begin{eqnarray}
\mathcal{L}^{q}_{\phi} & = & \frac{1}{2}\left[\partial_{\mu}\phi^{T}\partial^{\mu}\phi+B_{\mu}B^{\mu}\phi^{T}\phi\right.\nonumber\\
&&\left.+i(\phi^{T}B_{\mu}\partial^{\mu}\phi-\partial_{\mu}\phi^{T}B^{\mu}\phi)\right],
\end{eqnarray}

\begin{eqnarray}\label{lagramus}
\mathcal{L}^{q}_{A_{\mu}} & = &
-\frac{1}{4}(\bar{D}_{\mu}A^{a}_{\nu}-\bar{D}_{\nu}A^{a}_{\mu})^{2}+\bar{\phi}^{\dagger}A_{\mu}A^{\mu}\bar{\phi},
\end{eqnarray}

\begin{eqnarray}
\mathcal{L}^{q}_{GF}=&&-\frac{1}{2\xi}[(\partial_{\mu}A^{\mu}_{a})^{2}+2C^{abc}\partial_{\mu}A^{\mu}_{a}\bar{A}^{b}_{\nu}A^{\nu}_{c}\nonumber\\
&&+(C^{abc}A_{\mu}^{b}A^{\mu}_{c})^{2}+\xi^{2}(\phi^{\dagger}T^{a}\bar{\phi})^{2}],
\end{eqnarray}

\begin{eqnarray}
\mathcal{L}^{q}_{ghost}&=&-\bar{D}_{\mu}\eta^{*}_{a}\bar{D}^{\mu}\eta_{a}+\xi
g^{2}\eta^{*}_{a}\bar{\phi}^{T}T^{a}T^{b}\bar{\phi}\eta_{b}.
\end{eqnarray}

Because of the Goldstone theorem, the number of scalar bosons that
acquire a gauge dependent mass and the number of massive gauge
fields should be the same as the number of spontaneously broken
symmetries. If we define

\begin{equation}
(M_{A}^{ab})^{2}=\bar{\phi}^{T}T^{a}T^{b}\bar{\phi},
\end{equation}

\noindent so that

\begin{equation}
M_{A}^{2}=\frac{g^{2}\nu^{2}}{4},
\end{equation}

\noindent the masses of the fields involved are given by

\begin{eqnarray}
m_{\phi_{1,2}}^{2}&=&m^{2}+\lambda\nu^{2}+\xi M_{A}^{2}\equiv m^{2}_{1},\nonumber\\
m_{\phi_{3}}^{2}&=&m^{2}+3\lambda\nu^{2}\equiv m^{2}_{3},\nonumber\\
m_{A_{1,2}}^{2}&=&M_{A}^{2},\nonumber\\
m_{A_{3}}^{2}&=&0,\nonumber\\
m_{\eta_{1,2}}^{2}&=&\xi M_{A}^{2},\nonumber\\
m_{\eta_{3}}^{2}&=&0.
\end{eqnarray}

Choosing $\xi=1$, and writing the Lagrangian in the form

\begin{eqnarray}
\int_{0}^{\beta}\hspace{-0.1 cm}{\textrm{d}\tau}\int \textrm{d}^3
x\mathcal{L}_{q}=\hspace{-0.1 cm}&-&\hspace{-0.1 cm}\frac{1}{2}\int
\hspace{-0.1 cm}\textrm{d}\bar{x}'\hspace{-0.1 cm}\int
\hspace{-0.1 cm}\textrm{d}\bar{x}\phi^a(\bar{x}')A_{ab}(\bar{x}',\bar{x})\phi^b(\bar{x})\nonumber\\
\hspace{-0.1 cm}&-&\hspace{-0.1 cm}\frac{1}{2}\int\hspace{-0.1 cm}
\textrm{d}\bar{x}'\hspace{-0.1 cm}\int\hspace{-0.1 cm}
\textrm{d}\bar{x}A^a_{\mu}(\bar{x}')B^{\mu\nu}_{ab}(\bar{x}',\bar{x})A^b_{\nu}(\bar{x})\nonumber\\
\hspace{-0.1 cm}&-&\hspace{-0.1 cm}\int \hspace{-0.1
cm}\textrm{d}\bar{x}'\hspace{-0.1 cm}\int\hspace{-0.1 cm}
\textrm{d}\bar{x}\omega^{*}(\bar{x}')C(\bar{x}',\bar{x})\omega(\bar{x}),
\end{eqnarray}

\noindent we have that the thermodynamical effective potential is
given by

\begin{equation}
-\beta\int \textrm{d}^3x \hspace{0.1
cm}\Omega_{eff}=-\frac{1}{2}\textrm{Tr}\ln\textbf{A}-\frac{1}{2}\textrm{Tr}\ln\textbf{B}+\textrm{Tr}\ln\textbf{C}.
\end{equation}

A straightforward calculation leads us to the several thermal
contributions from the different fields. For the $\phi_{3}$ boson we
have

\begin{equation}\label{uno}
\Omega^{\beta}_{\phi_{3}}=\frac{1}{\beta}\int d^{3}k\ln
(1-e^{-\beta(\sqrt{{\bf k}^{2}+m^{2}_{3}})}),
\end{equation}

and for $\phi_{1}$ and $\phi_{2}$

\begin{eqnarray}\label{dos}
\Omega^{\beta}_{\phi_{1,2}}=\frac{1}{\beta}\int \hspace{-0.15
cm}&d^{3}k&\hspace{-0.15 cm}\left[ \ln(1-e^{-\beta (\sqrt{{\bf
k}^{2}+m^{2}_{1}}
+\frac{\mu}{2})})\right. \nonumber\\
\hspace{-0.15 cm}&+&\left. \ln(1-e^{-\beta (\sqrt{{\bf
k}^{2}+m^{2}_{1}}-\frac{\mu}{2})})\right].
\end{eqnarray}

Notice that $\mathcal{L}^{q}_{A_{\mu}}$ is very similar to that
calculated in (\ref{amu}), but now two of the gauge fields are
massive, because of the $\bar{\phi}A_{\mu}A^{\mu}\phi$ term in
(\ref{lagramus}). The contribution of these fields is

\begin{widetext}

\begin{equation}
\Omega^{\beta}_{A_{\mu}}=\frac{1}{\beta}\int{d^{3}k\left(4\ln{\left[(1-e^{-\beta(\sqrt{{\bf
k}^{2}+M^{2}_{A}}+\mu)})(1-e^{-\beta(\sqrt{{\bf
k}^{2}+M^{2}_{A}}-\mu)})\right]}+4\ln(1-e^{-\beta |{\bf
k}|})\right)}.
\end{equation}

When we add the contribution of the Faddeev-Popov Lagrangian we obtain

\begin{equation}\label{tres}
\Omega^{\beta}_{A_{\mu},\eta}=\frac{1}{\beta}\int{d^{3}k\left(2\ln{\left[(1-e^{-\beta(\sqrt{{\bf
k}^{2}+M^{2}_{A}}+\mu)})(1-e^{-\beta(\sqrt{{\bf
k}^{2}+M^{2}_{A}}-\mu)})\right]}+2\ln(1-e^{-\beta |{\bf
k}|})\right)}.
\end{equation}

\end{widetext}

Notice that the number of degrees of freedom is the expected one. If
we choose $\nu^{2}=-m^{2}/\lambda$ we recover the usual Higgs-Kibble
mechanism. In the presence of chemical potential, however, we have a
lesser number of Goldstone bosons, as was shown in \cite{miransky}.
Our calculation confirms this picture.

The treatment of the fermion fields is equivalent to the usual
procedure, where the chemical potentials appear as external zero
component gauge fields forming a new covariant derivative. This a
consequence of the fact that the conserved fermionic charge does not
depend on the derivatives of the fields, i.e. there is no need to
integrate over the conjugate momenta to pass from the Hamiltonian
picture to the Lagrangian formalism. For the fermion fields we have

\begin{eqnarray}
\mathcal{L}_{\psi}= i\bar{\psi}(\partial \hspace{-0.2
cm}/-igA\hspace{-0.2 cm}/-igB\hspace{-0.23 cm}/\hspace{0.1
cm})\psi+m^{2}_{\psi}\bar{\psi}\psi,
\end{eqnarray}

\noindent where $m^{2}_{\psi}$ is the mass of the fermions due to
the Higgs-Kibble mechanism. The calculation of their contribution to
the thermodynamical potential is straightforward. We found the well
known result

\begin{eqnarray}\label{cuatro}
\Omega^{\beta}_{\psi}=-\frac{1}{\beta}\int \hspace{-0.2
cm}&d^{3}k&\hspace{-0.2 cm}\left[ 2\ln(1+e^{-\beta (\sqrt{{\bf
k}^{2}+m^{2}_{\psi}}
+\frac{\mu}{2})})\right. \nonumber\\
&+&\left. 2\ln(1+e^{-\beta (\sqrt{{\bf
k}^{2}+m^{2}_{\psi}}-\frac{\mu}{2})})\right].
\end{eqnarray}

The final one loop effective potential in a $SU(2)$ gauge theory
with scalars and fermions will be given by the sum of equations
(\ref{uno}), (\ref{dos}), (\ref{tres}) and (\ref{cuatro}).

In this letter we have shown that the gauge fixing condition that
enables to diagonalize the effective potential for a system
including gauge, scalar and/or fermion fields emerges naturally from
the description based on the BFM method.

This gauge fixing condition has been only explored for small gauge
field configurations. The analysis of the existence of Gribov copies
will be carried on in a future work.

\section*{ACKNOWLEDGMENTS}

The authors would like to thank financial support from
 FONDECYT under grant 1051067.

\end{document}